\documentclass[aps,pra,onecolumn,superscriptaddress,groupedaddress]{revtex4}  
\usepackage{graphicx}  
\usepackage{dcolumn}   
\usepackage{bm}        
\usepackage{amsthm}
\usepackage{amsmath}
\usepackage{amssymb}
\usepackage{amsfonts}
\usepackage{amscd}
\usepackage{enumerate}
\usepackage{bm}
\usepackage{latexsym}
\usepackage{color}
\usepackage{hyperref}


\def\;{{\hspace{0.3ex};\hspace{0.5ex}}}
\def\,{{\hspace{0,3ex},\hspace{0.5ex}}}
\def\({{\hspace{1.2ex}(}}

\def\QED{\mbox{\rule[0pt]{1.5ex}{1.5ex}}}

\def\endproof{\hspace*{\fill}~\QED\par\endtrivlist\unskip}

\newtheorem{definition}{Definition}
\newtheorem{proposition}[definition]{Proposition}
\newtheorem{lemma}[definition]{Lemma}

\newtheorem{theorem}[definition]{Theorem}
\newtheorem{corollary}[definition]{Corollary}
\newtheorem{conjecture}[definition]{Conjecture}

\newtheorem{remark}[definition]{Remark}
\newtheorem{example}[definition]{Example}
\newtheorem{question}[definition]{Question}

\newcommand{\nc}{\newcommand}

 \nc{\bbA}{\mathbb{A}} \nc{\bbB}{\mathbb{B}} \nc{\bbC}{\mathbb{C}}
 \nc{\bbD}{\mathbb{D}} \nc{\bbE}{\mathbb{E}} \nc{\bbF}{\mathbb{F}}
 \nc{\bbG}{\mathbb{G}} \nc{\bbH}{\mathbb{H}} \nc{\bbI}{\mathbb{I}}
 \nc{\bbJ}{\mathbb{J}} \nc{\bbK}{\mathbb{K}} \nc{\bbL}{\mathbb{L}}
 \nc{\bbM}{\mathbb{M}} \nc{\bbN}{\mathbb{N}} \nc{\bbO}{\mathbb{O}}
 \nc{\bbP}{\mathbb{P}} \nc{\bbQ}{\mathbb{Q}} \nc{\bbR}{\mathbb{R}}
 \nc{\bbS}{\mathbb{S}} \nc{\bbT}{\mathbb{T}} \nc{\bbU}{\mathbb{U}}
 \nc{\bbV}{\mathbb{V}} \nc{\bbW}{\mathbb{W}} \nc{\bbX}{\mathbb{X}}
 \nc{\bbZ}{\mathbb{Z}}

\def\min{\mathop{\rm min}}
\def\diag{\mathop{\rm diag}}

\nc{\cA}{{\cal A}} \nc{\cB}{{\cal B}} \nc{\cC}{{\cal C}}
\nc{\cD}{{\cal D}} \nc{\cE}{{\cal E}} \nc{\cF}{{\cal F}}
\nc{\cG}{{\cal G}} \nc{\cH}{{\cal H}} \nc{\cI}{{\cal I}}
\nc{\cJ}{{\cal J}} \nc{\cK}{{\cal K}} \nc{\cL}{{\cal L}}
\nc{\cM}{{\cal M}} \nc{\cN}{{\cal N}} \nc{\cO}{{\cal O}}
\nc{\cP}{{\cal P}} \nc{\cQ}{{\cal Q}} \nc{\cR}{{\cal R}}
\nc{\cS}{{\cal S}} \nc{\cT}{{\cal T}} \nc{\cU}{{\cal U}}
\nc{\cV}{{\cal V}} \nc{\cW}{{\cal W}} \nc{\cX}{{\cal X}}
\nc{\cZ}{{\cal Z}}

\def\a{\alpha}
\def\b{\beta}
\def\g{\gamma}
\def\d{\delta}
\def\e{\epsilon}

\def\t{\theta}

\def\x{\xi}
\def\p{\pi}

\def\u{\upsilon}

\def\o{\omega}

\def\dg{\dagger}

\def\op{\oplus}
\def\ox{\otimes}

\newcommand{\bra}[1]{\langle#1|}
\newcommand{\ket}[1]{|#1\rangle}
\newcommand{\proj}[1]{| #1\rangle\!\langle #1 |}

\newcommand{\braket}[2]{\langle#1|#2\rangle}
\newcommand{\abs}[1]{|#1|}

\nc{\U}{\mathrm{U}}

\def\bcj{\begin{conjecture}}
\def\ecj{\end{conjecture}}
\def\bcr{\begin{corollary}}
\def\ecr{\end{corollary}}
\def\bd{\begin{definition}}
\def\ed{\end{definition}}
\def\bea{\begin{eqnarray}}
\def\eea{\end{eqnarray}}
\def\bem{\begin{enumerate}}
\def\eem{\end{enumerate}}
\def\bex{\begin{example}}
\def\eex{\end{example}}
\def\bim{\begin{itemize}}
\def\eim{\end{itemize}}
\def\bl{\begin{lemma}}
\def\el{\end{lemma}}
\def\bma{\begin{bmatrix}}
\def\ema{\end{bmatrix}}
\def\bpf{\begin{proof}}
\def\epf{\end{proof}}
\def\bpp{\begin{proposition}}
\def\epp{\end{proposition}}
\def\bqu{\begin{question}}
\def\equ{\end{question}}
\def\br{\begin{remark}}
\def\er{\end{remark}}
\def\bt{\begin{theorem}}
\def\et{\end{theorem}}

\begin{document}


\title{Mutually unbiased bases in dimension six containing a product-vector basis}

\author{Lin Chen}\email{linchen@buaa.edu.cn}
\affiliation{School of Mathematics and Systems Science, Beihang University, Beijing 100191, China}
\affiliation{International Research Institute for Multidisciplinary Science, Beihang University, Beijing 100191, China}
\author{Li Yu}\email{yupapers@sina.com}
\affiliation{Department of Physics, Hangzhou Normal University, Hangzhou, Zhejiang 310036, China}
\date{\today}

\begin{abstract}
Excluding the existence of four MUBs in $\bbC^6$ is an open problem in quantum information.
We investigate the number of product vectors in the set of four mutually unbiased bases (MUBs) in dimension six, by assuming that the set exists and contains a product-vector basis. 
We show that in most cases the number of product vectors in each of the remaining three MUBs is at most two. We further construct the exceptional case in which the three MUBs respectively contain at most three, two and two product vectors. We also investigate the number of vectors mutually unbiased to an orthonormal basis. 

\end{abstract}




\maketitle


\section{Introduction}

Deciding the maximum number of mutually unbiased bases (MUBs) in $\bbC^6$ is a well-known open problem in quantum information theory. The study of MUBs has various applications in quantum cryptography and quantum tomography, and in fundamental problems such as the construction of Wigner functions. It has been shown that there are three MUBs in $\bbC^6$, and it has been widely conjectured that four MUBs in $\bbC^6$ do not exist. Recent progress on MUBs and the conjecture can be found in \cite{Boykin05,bw08,bw09,jmm09,mub09,bw10,deb10,wpz11,mw12ijqi,mw12jpa135307,rle11,Sz12,Goyeneche13,mw12jpa102001,mb15,mpw16,cy17,gmm17}. 
We adopt the notation that a unitary matrix corresponds to an orthonormal basis consisting of the column vectors of the matrix. In particular we refer to \emph{identity} as the identity matrix corresponding to the computational basis. We have investigated in \cite{cy17} the conjecture in terms of the product vectors and Schmidt rank of the unitary matrices corresponding to the bases. We review the main result of \cite{cy17} as follows.
\bl
\label{le:mub6}
Suppose the set of four MUBs in $\bbC^6$ exists. If it contains the identity, then
\\
(i) any other MUB in the set contains at most two product column vectors;
\\
(ii) the other three MUBs in the set contains totally at most six product column vectors.
\el
In this paper we extend the result by replacing the identity by an arbitrary product-vector basis. The latter has been classified into three sets, namely $\cP_1,\cP_2$ and $\cP_3$ in Lemma \ref{le:la} (xxi). We shall show in Proposition \ref{pp:prod} (ii) that if one of the product-vector basis and the set of product states in another MUB is not from $\cP_1$ up to local unitaries then the claims (i) and (ii) in Lemma \ref{le:mub6} both hold. Otherwise, namely if the product-vector basis and the set of product states in another MUB are both from $\cP_1$ up to local unitaries, then the number of product states in the MUB is at most three, as we show in Proposition \ref{pp:prod} (i). If the number is exactly three, then we show in Proposition \ref{pp:6+3+2+2} (i) that the remaining two non-product-vector MUBs of the four MUBs in $\bbC^6$ each has at most two product vectors. So the number of product vectors in such a set of four MUBs is at most $6+3+2+2=13$. This does not mean that there cannot be four MUBs with more product vectors. We have not excluded the possibility that there are $4$ product vectors in each of the four bases. On the other hand, if a basis contains $5$ product vectors, then it contains $6$ product vectors, according to \cite[Lemma 6(xx)]{cy17}. We investigate the expressions of product states and the remaining entangled states in the MUBs in Proposition \ref{pp:6+3+2+2} (ii) and (iii).

To obtain the above results, we start by introducing preliminary Lemma \ref{le:la}. Then we investigate the number of vectors mutually unbiased to an orthonormal basis. It is a more general problem than the existence of MUBs. We list the vectors when $d=2,3$ in Lemma \ref{le:mubsinkhorn}, and prove some statements about product vectors unbiased to some other product states when $d=6$. Then we reiterate some results on MUBs and complex Hadamard matrices in Lemma \ref{le:mub}.

It is known that two quantum states in $\bbC^d$ are MU when their inner product has modulus ${1\over\sqrt d}$. Two orthonormal basis in $\bbC^d$ are MU if their elements are all MU.
We can similarly define $n$ orthonormal basis in $\bbC^d$ as $n$ MUBs in $\bbC^d$, if any two of them are MU. 
It is easy to see that if the first MUB is the identity matrix then all other MUBs must be complex Hadamard matrices (CHM), i.e., unitary matrices whose entries all have the same modulus $1/\sqrt d$. From now on we regard any order-six CHM as a $2\times3$ bipartite unitary operation. The latter has been recently extensively studied in terms of the entangling power, assisted entangling  power of bipartite unitaries and the relation to controlled unitary operations \cite{cy13,cy14,cy14ap,cy15,cy16,cy16b,cy17}. The first two quantities quantitatively characterize the maximum amount of entanglement increase when the input states are respectively a product state and arbitrary pure states. The maximum amount of entanglement increase over all input states is a lower bound of the entanglement cost for implementing bipartite unitaries under local operations and classical communications. Further, controlled unitary operations such as CNOT gates are fundamental ingredients in quantum computing.

The rest of the paper is organized as follows. We introduce the notations and preliminary results in Sec. \ref{sec:pre}. They include the equivalent MUBs, the MUB trio and linear algebra. We investigate the number of vectors mutually unbiased to an orthonormal basis in Sec. \ref{sec:min}. We construct our main results in Sec. \ref{sec:mub}. Finally we conclude in Sec. \ref{sec:con}.

\section{Preliminaries}
\label{sec:pre}

In this section we introduce the notations and preliminary facts used in the paper. Let $I_d$ (abbreviated as $I$ when $d$ is known) denote the order-$d$ identity matrix. Let $\ket{i,j},i=1,\cdots,d_A$, $j=1,\cdots,d_B$ be the computational-basis states of the bipartite Hilbert space $\cH=\cH_A\ox\cH_B=\bbC^{d_A}\ox\bbC^{d_B}$. We shall refer to $\ket{a}$ and $\ket{a^\perp}$ as two orthonormal states. If $S$ is a subspace then $S^\perp$ is the orthogonal subspace of $S$. For $d=pq$ with $p,q>1$, the basis of $\bbC^d$ consisting of product vectors in $\bbC^p\ox\bbC^q$ is called a product-vector basis.  We say that $n$ unitary matrices form $n$ product-vector MUBs when the column vectors of these matrices are all product vectors and they form $n$ MUBs.  Next, a square matrix $C$ is a direct-product matrix if $C=F\ox G$ where $F$ and $G$ are square matrices of order greater than one. The \textit{subunitary matrix} is matrix proportional to a unitary matrix. We review the following definition from \cite{cy17}.
\bd
\label{df:mub}
(i) Let $U_1,\cdots,U_n$ be $n$ unitary matrices of order $d$. They form $n$ MUBs if and only if for an arbitrary unitary matrix $X$, and arbitrary complex permutation matrices $P_1,\cdots,P_n$, the $n$ matrices $XU_1P_1,\cdots,XU_nP_n$ form $n$ MUBs.
In this case we say that $U_1,\cdots,U_n$ and $XU_1P_1,\cdots,XU_nP_n$ are unitarily equivalent MUBs. Furthermore they are locally unitarily (LU) equivalent MUBs when $X$ is a direct-product matrix.

Let $U_1,\cdots,U_n$ be product-vector MUBs such that $U_j=(\cdots,\ket{a_{jk},b_{jk}},\cdots)$ where $\ket{a_{jk}}\in\bbC^p$ and $\ket{b_{jk}}\in\bbC^q$. Let
$U_j^{\Gamma_A}$ and $U_j^{\Gamma_B}$ both denote $U_j$ except that $\ket{a_j}$ and $\ket{b_j}$ are respectively replaced by their complex conjugates. Then we say that any two of the following four sets
$
U_1,\cdots,U_n,
U_1^{\Gamma_A},\cdots,U_n^{\Gamma_A},
U_1^{\Gamma_B},\cdots,U_n^{\Gamma_B},
XU_1P_1,\cdots,XU_nP_n,
$
are LU-equivalent product-vector MUBs, where $X$ is a direct-product matrix.
\\
(ii) Let $U,V$ and $W$ be three CHMs of order six. The existence of four MUBs in $\bbC^6$ is equivalent to ask whether $I,U,V$ and $W$ can form four MUBs, i.e., whether $U^\dg V$, $V^\dg W$ and $W^\dg U$ are still CHMs. If they do, then we denote the set of $U,V$ and $W$ as an MUB trio.
\\
(iii) We say that two CHMs $X$ and $Y$ are equivalent when there exist two complex permutation matrices $C$ and $D$ such that $X=CYD$. For simplicity we refer to $X$ as $Y$ up to equivalence. The equivalence class of $X$ is the set of all CHMs which are equivalent to $X$.
\\(iv) In (iii), we say that $X$ and $Y$ are locally equivalent when $C$ is a direct-product matrix.
\ed

The following result on linear algebra is from \cite[Lemma 6]{cy17}.
\bl
\label{le:la}
(i) Suppose an orthonormal basis in $\bbC^6$ contains $k$ product states with $k=0,1,\cdots,5$. Then the remaining $6-k$ states in the basis span a subspace spanned by orthogonal product vectors.

(ii) Any product-vector basis in $\bbC^2\ox\bbC^3$ is LU equivalent to one of the following three sets of orthonormal bases,
\bea
\label{eq:cp1}
\cP_1:=\{\ket{0,0},\ket{0,1},\ket{0,2},
\ket{1,a_0},\ket{1,a_1},\ket{1,a_2}\};
\\\label{eq:cp2}
\cP_2:=\{
\ket{0,0},\ket{0,1},
\ket{1,b},\ket{1,b^\perp},
\ket{c,2},\ket{c^\perp,2}
\};
\\\label{eq:cp3}
\cP_3:=\{
\ket{0,0},\ket{1,0},
\ket{d,1},\ket{d^\perp,1},
\ket{e,2},\ket{e^\perp,2}\}.
\eea
Here $\{\ket{a_i}\}$ is an orthonormal basis in $\bbC^3$, $\{\ket{b},\ket{b^\perp}\}$ is an orthonormal basis in $\bbC^2$, the first row and column of the matrix $[\ket{a_0},\ket{a_1},\ket{a_2}]$ are all $1/\sqrt3$, $\ket{b},\ket{b^\perp}$, $\ket{c},\ket{c^\perp}$, $\ket{d},\ket{d^\perp}$ are all real, the first elements of $\ket{e}$ and $\ket{e^\perp}$ are both real.

(iii) Let $\abs{a}^2+\abs{b}^2=1$, $\ket{v}$, $\ket{w}\in\bbC^3$, and $a\ket{v}+b\ket{w}$ be of elements of modulus $1/\sqrt3$. Then $ab=0$ or $[\ket{v},\ket{w}]$ is a matrix of size $3\times2$ equivalent to a real matrix.
\qed
\el

Evidently the computational basis $\{\ket{0,0},\ket{0,1},\ket{0,2},\ket{1,0},\ket{1,1},\ket{1,2}\}=\cP_1\cap\cP_3=\cP_1\cap\cP_2\cap\cP_3$. 
We claim that up to phases,
\bea
\label{eq:cb}
\{
\ket{0,0},\ket{0,1},
\ket{1,b},\ket{1,b^\perp},
\ket{0,2},\ket{1,2}
\}
&=&
\cP_1\cap
\{(W\ox X)\cP_2,
~~
\forall W, X\},
\\
\label{eq:cb1}
\{\ket{0,0},\ket{0,1},\ket{0,2},\ket{1,0},\ket{1,1},\ket{1,2}\}
&=&
\cP_1\cap
\{(U\ox V)\cP_3,~~
\forall U,V\}
\notag\\
&=&\cP_1\cap\{(W\ox X)\cP_2\cap(U\ox V)\cP_3,
~~
\forall W, X, U, V\},
\eea
where $W,U$ are order-two unitary matrices and $X,V$ are order-three unitary matrices. To prove 
\eqref{eq:cb}, we can see that the lhs of \eqref{eq:cb} belongs to the rhs of \eqref{eq:cb} by assuming $W\ket{c}=\ket{0}$ and $X$ as the identity matrix. Next suppose $x\in \cP_1\cap(W\ox X)\cP_2$. We obtain that $W\ket{c}=\ket{0}$ or $\ket{1}$. So $x$ belongs to the lhs of \eqref{eq:cb}. We have proved that the lhs and rhs of \eqref{eq:cb} are the same. One can similarly prove \eqref{eq:cb1}. 

The relation \eqref{eq:cb1} shows that the computational basis is the unique element in the intersection of the three orthonormal product states in $\bbC^2\ox\bbC^3$.
The computational basis corresponds to the identity matrix in the four MUBs in $\bbC^6$, and we have studied the case in \cite{cy17}. So the case is the basis of studying the four MUBs in $\bbC^6$ containing a product-vector basis, as we will see in Sec. \ref{sec:mub}.

\section{The number of vectors mutually unbiased to an orthonormal basis}
\label{sec:min}

We say that a matrix is in the \textit{dephased form} when all elements in the first row and first column of the matrix are  real and nonnegative. Evidently every CHM is equivalent to another CHM in the dephased form. We say a vector is dephased if it is a zero vector or if its first nonzero element is real and positive. For any order-$d$ unitary $U$, we denote \textit{an MU vector} of $U$ as a dephased normalized vector unbiased to all column vectors of both $I_d$ and $U$. Let $N_v(U)$ denote the number of such vectors. Such vectors provide examples of the so-called zero noise, zero disturbance (ZNZD) states for two orthonormal bases  consisting of the column vectors of $I_d$ and $U$ \cite{kjr14}.
The $N_v$ is not an invariant under local unitary operations. A counterexample is as follows. Let $U=I_2 \ox F_3$, where $F_3$ is the order-three Fourier matrix. Then $N_v(U)=\infty$, though there is an order-two unitary $X$ such that $N_v[(X \ox I_3)U]$ is finite. The problem of finding MU vectors is a more general problem than constructing MUBs, because sometimes the found MU vectors do not form an orthonormal basis or some set of MUBs. Finding four MUBs in $\bbC^6$ requires to find out $18$ unit vectors MU to a given orthonormal basis, and these 18 vectors need to form three MUBs. To study $N_v(U)$ we construct a preliminary lemma.

\bl\label{le:mubsinkhorn}
Let $d$ be an integer greater than $1$.
\\
(i) For any two orthonormal bases in $\bbC^d$, there is a normalized vector MU to both bases.
Equivalently, for any unitary matrix $U$ of order $d$, we have $N_v(U)\ge1$.
\\
(ii) For any two MUBs in $\bbC^d$, there is a normalized vector unbiased to both MUBs.
Equivalently, for any CHM $U$ of order $d$, we have $N_v(U)\ge1$.
\\
(iii) Suppose $d=2$ and the $U$ in (ii) is ${1\over\sqrt2}\bma 1 & 1 \\ 1 & -1\ema$. Then $N_v(U)=2$, and the two vectors are $(1,i)/\sqrt2$ and $(1,-i)/\sqrt2$. 
\\
(iv) Suppose $d=3$ and the $U$ in (ii) is ${1\over\sqrt3}\bma 1 & 1 & 1 \\ 1 & \o & \o^2 \\ 1&\o^2&\o\ema$. Then  $N_v(U)=6$, and the six vectors are the column vectors in the  matrix
$
{1\over\sqrt3}
\bma
1 & 1 & 1 & 1 & 1 & 1\\
\o & \o^2 & 1 & \o^2 & \o & 1 \\
\o & 1 & \o^2 & \o^2 & 1 & \o\\
\ema.
$
\\
(v) Suppose two orthogonal product vectors $\ket{a,b},\ket{a,b^\perp}$ are MU to another two orthogonal product vectors $\ket{c,d},\ket{c,d^\perp}$, where $\ket{b},\ket{b^\perp}$, $\ket{d},\ket{d^\perp}$ are 3-dimensional vectors of elements of modulus $1/\sqrt3$. Then $\ket{a}$ and $\ket{c}$ are MU, and $\ket{b},\ket{b^\perp}$ and $\ket{d},\ket{d^\perp}$ are also MU. Further if $\ket{b},\ket{b^\perp}$ are two column vectors with the form $\bma1 \\ \o^m \\ \o^n\ema$ with some integers $m,n$, then so are $\ket{d},\ket{d^\perp}$. 
\el
\bpf
Assertion (i) and (ii) have been proved in \cite{cy17}. Assertion (iii) and (iv) follow from Eqs. (2.6) and (2.10) in \cite{mub09}.

It remains to prove (v). We can find a complex permutation matrix $P$ such that $P\ket{b}={1\over\sqrt3}\bma1\\1\\1\ema$ and $P\ket{b^\perp}={1\over\sqrt3}\bma1\\\o\\\o^2\ema$. They are MU to the two orthogonal product vectors $P\ket{d}\propto{1\over\sqrt3}\bma1\\x_1\\y_1\ema$ and $P\ket{d^\perp}\propto{1\over\sqrt3}\bma1\\x_2\\y_2\ema$ where $\abs{x_j}=\abs{y_j}=1$. Suppose $\ket{a},\ket{c}\in\bbC^n$.
The hypothesis implies that 
\bea
\label{eq:1+x1y1}
\abs{1+x_1+y_1}=\abs{1+x_2+y_2}=
\abs{1+x_1\o^2+y_1\o}=\abs{1+x_2\o^2+y_2\o}=
{\sqrt {3}\over\abs{\braket{a}{c}} \sqrt n}.
\eea 
The orthogonality implies $1+x_1x_2^*+y_1y_2^*=0$. So $(x_1x_2^*,y_1y_2^*)=(\o,\o^2)$ or $(\o^2,\o)$. In either case, \eqref{eq:1+x1y1} implies that the two vectors ${1\over\sqrt3}\bma1\\x_1\\y_1\ema$ and ${1\over\sqrt3}\bma1\\x_2\\y_2\ema$ are both MU to the column vectors of ${1\over\sqrt3}
\bma                    
1 & 1 & 1 \\
1 & \o& \o^2 \\
1 & \o^2 & \o \\                    
 \ema$. It follows from Lemma \ref{le:mubsinkhorn} (iii) that the two vectors are from the column vectors of $
{1\over\sqrt3}
\bma
1 & 1 & 1 & 1 & 1 & 1\\
\o & \o^2 & 1 & \o^2 & \o & 1 \\
\o & 1 & \o^2 & \o^2 & 1 & \o\\
\ema.
$
So \eqref{eq:1+x1y1} implies that $\abs{\braket{a}{c}}=1/\sqrt n$. 
We have proved the first assertion of (v). The second assertion follows from the use of $P$ in the above proof. 
This completes the proof.
\epf

We claim that there exist order-$d$ unitary matrices $U$, such that $N_v(U)=\infty$ or $2$. An example for the former claim is any non-identity permutation matrix $U$, and an example for the latter claim is $U=\bma \cos\a & \sin\a \\ \sin\a & -\cos\a \ema$ where $\a\in(0,\p/2)$. On the other hand if $U$ is a CHM, it is known that $N_v(U)$ is finite when $d=2,3$ and $5$, and infinite when $d=4$ \cite{cy17}.
More results on the minimum number of MUBs and $N_v(U)$ can be found in \cite[Table 1]{gmm17}. Lemma \ref{le:mubsinkhorn} will be used in the proof of Proposition \ref{pp:prod}. The following lemma has been proved in \cite{cy17}.

\bl
\label{le:mub}
(i) If a normalized vector is MU to $d-1$ vectors in an orthonormal basis in $\bbC^d$, then it is also MU to the $d$'th vector in the basis.
\\
(ii) An order-six CHM is a member of some MUB trio if and only if so is its adjoint matrix, if and only if so is its complex conjugate, and if and only if so is its transpose.
\\
(ii.a) Let $k$ be a positive integer at most three. Then $k$ order-six CHMs are the members of some MUB trio if and only if so are their complex conjugate.
\\
(iii) The product state $\ket{a,b}$ in $\bbC^d$ is MU to an orthogonal product-vector basis $\{\ket{x_i,y_i}\}_{i=1,\cdots,d}$ if and only if $\ket{a}$ is MU to $\{\ket{x_i}_{i=1,\cdots,d}$ and $\ket{b}$ is MU to $\{\ket{y_i}\}_{i=1,\cdots,d}$.
\\
(iv) Any set of three product-vector MUBs in the space $\bbC^2\ox\bbC^3$ is LU equivalent to either
$\cT_0:=\{\ket{a_j,d_k},\ket{b_j,e_k},\ket{c_j,f_k}\}
$
or
$
\cT_1:=\{\ket{a_j,d_k},\ket{b_j,e_k},\ket{c_0,f_k},\ket{c_1,g_k}\},$
where $\{\ket{a_j}\},\{\ket{b_j}\}$ and $\{\ket{c_j}\}$ is a complete set of MUBs in $\bbC^2$, and $\{\ket{d_j}\},\{\ket{e_j}\},\{\ket{f_j}\}$ and $\{\ket{g_j}\}$ is a complete set of MUBs in $\bbC^3$.
\\
(v) Any set of three product-vector MUBs in the space $\bbC^2\ox\bbC^3$ is not MU to a single state.
\\
(vi) Any set of four MUBs in $\bbC^6$ contains at most one product-vector basis. Equivalently, any two of four MUBs in $\bbC^6$ contain at most ten product vectors.
\\
(vii) Any MUB trio contains none of the order-six CHMs
$Y_1,\cdots,Y_7$ where
\bem
\item
$Y_1$ contains an order-three subunitary matrix.
\item
$Y_2$ contains a submatrix of size $3\times2$ and rank one.
\item
$Y_3$ contains an order-three submatrix whose one column vector is orthogonal to the other two column vectors.
\item
three column vectors of $Y_4$ are product vectors.
\item
$Y_5$ contains an order-three singular submatrix.
\item
$Y_6$ contains a real submatrix of size $3\times2$.
\item
two column vectors of $Y_7$ are product vectors $\ket{a,b}$ and $\ket{a,c}$.
\eem
\qed
\el

\section{Mutually unbiased bases containing a product-vector basis}
\label{sec:mub}

In this section we present the main results of this paper, namely Proposition \ref{pp:prod} and \ref{pp:6+3+2+2}. In Lemma \ref{le:mub} (vi), we have investigated when two of four MUBs in $\bbC^6$ have at most 10 product column vectors. We hope to further decrease the number. This is a problem different from
Lemma \ref{le:mubsinkhorn}, in which the two MUBs may not belong to a set of four MUBs. The motivation of the problem is as follows. We have met many four MUBs in which an MUB consists of product column vectors. The product column vectors in other three MUBs may decide the structure of the four MUBs or their existence. An approach to the problem is assuming that up to local unitaries, an MUB is from $\cP_1,\cP_2$ and $\cP_3$ in Lemma \ref{le:la} (ii).
If it corresponds to the identity matrix, each of the other three MUB contains at most two product vectors by Lemma \ref{le:mub6}. 
On the other hand, we have shown that the identity matrix is a subcase of the product-vector basis in Lemma \ref{le:la} (ii). So studying the MUBs under the assumption extends Lemma \ref{le:mub6} more generally helps understand the existence of four MUBs in $\bbC^6$. We begin by presenting the following two lemmas.

\bl
\label{le:2prod2}
Suppose there are four MUBs in $\bbC^6$, and the first one of them is $\cP_j$ for $j\in\{1,2,3\}$. Then any product state in other three MUBs has elements of modulus $1/\sqrt6$. That is, up to global phases the product state has the expression $(1,u)^T/\sqrt2 \ox (1,v,w)^T/\sqrt3$ where $\abs{u}=\abs{v}=\abs{w}=1$.

Suppose one of the other three MUBs contains exactly $n$ product states. We have
\\
(i) if $j=2$, then $n\le2$;
\\
(ii) if $j=3$, then $n\le2$.
\el
\bpf
Let the product state be $(a,b)^T\ox(c,d,e)^T$. It follows from Lemma \ref{le:la} (ii) and Lemma \ref{le:mub} (iii) that $\abs{a}=\abs{b}=1/\sqrt2$ and $\abs{c}=\abs{d}=\abs{e}=1/\sqrt3$. We have proved the first assertion. Next we prove the second assertion consisting of (i) and (ii). Suppose the second MUB of the four MUBs is the order-six unitary matrix $U$, and it contains exactly $n$ product states. Using the first assertion, we may assume that one of the $n$ product states is
\bea
\label{eq:1u}
(1,u)^T/\sqrt2 \ox (1,v,w)^T/\sqrt3,
\eea
where $\abs{u}=\abs{v}=\abs{w}=1$.

(i) It follows from Lemma \ref{le:la} (ii) that the first MUB is $\cP_2=\{
\ket{0,0},\ket{0,1},
\ket{1,b_0},\ket{1,b_1},
\ket{c_0,2},\ket{c_1,2}
\}$ with real orthonormal states $\ket{b_0},\ket{b_1}\in\bbC^2$ and real orthonormal states $\ket{c_0},\ket{c_1}\in\bbC^2$. If both of them are the basis $\ket{0},\ket{1}$ then $n\le2$ follows from Lemma \ref{le:mub6}.
Suppose $\ket{b_0},\ket{b_1}$ is not the basis $\ket{0}$, $\ket{1}$.
It follows from Lemma \ref{le:mub} (iii) and \eqref{eq:1u} that $\ket{b_0},\ket{b_1}$ are both MU to $(1,v,w)^T/\sqrt3$. Hence $v=i$ or $-i$.
We can assume that the upper left submatrix of $U$ of size $2\times n$ is $V={1\over\sqrt6}\bma 1 & \cdots & 1 \\ p_1 i & \cdots & p_n i \ema$ where $p_j=1$ or $-1$. Let $I_2\op W$ be an order-six unitary such that $(I_2\op W)\cP_2=I_6$, and the upper left submatrix of $(I_2\op W)U$ of size $2\times n$ is still $V$. Since $\cP_2$ and $U$ are two members of four MUBs, $(I_2\op W)U$ is a member of some MUB trio. We have $n\le2$ by Lemma \ref{le:mub} (ii) and the matrix $Y_6$ in (vii).

The remaining case is that $\ket{b_0},\ket{b_1}$ is the basis $\ket{0}$, $\ket{1}$, and $\ket{c_0},\ket{c_1}$ is not the basis $\ket{0}$, $\ket{1}$. It follows from Lemma \ref{le:mub} (iii) and \eqref{eq:1u}  that $\ket{c_0},\ket{c_1}$ are both MU to $(1,u)^T/\sqrt2$. Hence $u=i$ or $-i$. We can assume that the $2\times n$ submatrix formed by the first and fourth rows of $U$ is ${1\over\sqrt6}\bma 1 & \cdots & 1 \\ p_1 i & \cdots & p_n i \ema$ where $p_j=1$ or $-1$. Let $X=I_2\ox I_2 + W \ox \proj{2}$ be an order-six unitary such that $X\cP_2=I_6$. So $XU$ is a member of some MUB trio. We have $n\le2$ by Lemma \ref{le:mub} (ii) and the matrix $Y_6$ in (vii).

(ii) It follows from Lemma \ref{le:la} (ii) that the first MUB is $\cP_3=\{
\ket{0,0},\ket{1,0},
\ket{d_0,1},\ket{d_1,1},
\ket{e_0,2},\ket{e_1,2}\}$. Here $\{\ket{d_i}\}$ and $\{\ket{e_i}\}$ are all orthonormal bases in $\bbC^2$, $\{\ket{d_i}\}$ and the first elements of $\{\ket{e_i}\}$ are both real.
If both of them are the basis $\ket{0},\ket{1}$ then $n\le2$ follows from Lemma \ref{le:mub6}. If one of $\ket{d_0},\ket{d_1}$ and $\ket{e_0},\ket{e_1}$ is the basis $\ket{0}$, $\ket{1}$, then $\cP_3$ is locally equivalent to some $\cP_2$.
So $n\le2$ follows from (i). Suppose neither of $\ket{d_0},\ket{d_1}$ and $\ket{e_0},\ket{e_1}$ is the basis $\ket{0}$, $\ket{1}$. It follows from Lemma \ref{le:mub} (iii) and \eqref{eq:1u}  that $\ket{d_0},\ket{d_1}$ are both MU to $(1,u)^T/\sqrt2$. Hence $u=i$ or $-i$. Using \eqref{eq:1u},
we can assume that the $2\times n$ submatrix formed by the first and fourth rows of $U$ is $V={1\over\sqrt6}\bma 1 & \cdots & 1 \\ p_1 i & \cdots & p_n i \ema$ where $p_j=1$ or $-1$. Let $X=I_2\ox \proj{0} + W \ox \proj{1}+ W' \ox \proj{2}$ be an order-six unitary such that $X\cP_3=I_6$. So $XU$ is a member of some MUB trio, and the $2\times n$ submatrix formed by the first and fourth rows of $XU$ is $V$. We have $n\le2$ by Lemma \ref{le:mub} (ii) and the matrix $Y_6$ in (vii).
This completes the proof.
\epf

\bl
\label{le:2prod}
Suppose there are four MUBs in $\bbC^6$, the first one of them is $\cP_1$ and the second one of them contains exactly $n$ product states. We have
\\
(i) if the $n$ product states are from $\cP_1$ up to local unitaries then $n\le3$. Further if $n=3$ then the three product states are $\ket{0,0},\ket{0,1},\ket{1,a_0}$ up to local unitaries;
\\
(ii) if the $n$ product states are from $\cP_2$ or $\cP_3$ up to local unitaries then $n\le2$.
\el
\bpf
Eq. \eqref{eq:cp1} says that $\cP_1=\{\ket{0,0},\ket{0,1},\ket{0,2},
\ket{1,a_0},\ket{1,a_1},\ket{1,a_2}\}$, where $\ket{a_0},\ket{a_1}$ and $\ket{a_2}$ is an orthonormal basis in $\bbC^3$. Suppose the second one of the four MUBs is $U=\{\ket{w_0,x_0},\cdots,\ket{w_{n-1},x_{n-1}},\ket{0,y_n}+\ket{1,z_n},\cdots,\ket{0,y_5}+\ket{1,z_5}\}$.
Let $V$ be an order-three unitary matrix such that $V\ket{a_i}=\ket{i}$ for $i=0,1,2$. Then $(I_3\op V)U$ is a member of some MUB trio, and it is an order-six CHM. So any $\ket{w_j}$ with $j\le n-1$ is of elements of modulus $1/\sqrt2$, any $\ket{x_j}$ with $j\le n-1$ is of elements of modulus $1/\sqrt3$,
and any $\ket{y_j}$ with $n\le j \le 5$ is of elements of modulus $1/\sqrt6$.
The upper left submatrix of size $3\times n$ of $(I_3\op V)U$ is $X:=(w_{0,0}\ket{x_0},\cdots,w_{n-1,0}\ket{x_{n-1}})$ where $\abs{w_{j,0}}=1/\sqrt2$.
It follows from Lemma \ref{le:la} (i) that $\ket{0,y_n}+\ket{1,z_n},\cdots,\ket{0,y_5}+\ket{1,z_5}$ span a $(6-n)$-dimensional subspace spanned by orthogonal product vectors $\ket{w_n,x_n},\cdots,\ket{w_5,x_5}$. The states $\ket{0,y_j}+\ket{1,z_j}$ is entangled because $U$ contains exactly $n$ product states. Lemma \ref{le:la} (ii) implies that $Z:=\{\ket{w_0,x_0},\cdots,\ket{w_5,x_5}\}$ is from some $\cP_j$ up to local unitaries. So the states $\ket{x_0},\cdots,\ket{x_5}$ are equal to the 3-dimensional states in the product states of $\cP_j$ up to local unitaries.

Suppose $Z$ is from $\cP_2$ up to local unitaries.
If $n\ge3$ then $X$ contains three column vectors which are linearly dependent, or one of which is orthogonal to the other two. It is a contradiction with $Y_3$ and $Y_5$ in Lemma \ref{le:mub} (vii), because $X$ is a submatrix of $(I_3\op V)U$ which is a member of some MUB trio. Hence $n\le2$. One can similarly show that if $Z$ is from $\cP_3$ up to local unitaries then $n\le2$. So we have proved (ii).

It remains to prove (i). Suppose $Z$ is from $\cP_1$ up to local unitaries. Let $n\ge4$. Lemma \ref{le:mub} (vi) shows that $n=4$.
The argument for (ii) shows that $\ket{w_0}=\ket{w_1}=\ket{w_4}$, $\ket{w_2}=\ket{w_3}=\ket{w_5}$, and $\ket{x_0},\ket{x_1},\ket{x_4}$ and $\ket{x_2},\ket{x_3},\ket{x_5}$ are two orthonormal basis of $\bbC^3$.
Since $\ket{w_0},\cdots,\ket{w_3}$ are all of elements of modulus $1/\sqrt2$, so are $\ket{w_4}$ and $\ket{w_5}$. Since $\ket{x_0},\cdots,\ket{x_3}$ are all of elements of modulus $1/\sqrt3$, so are $\ket{x_4}$ and $\ket{x_5}$.
Recall that $\ket{0,y_4}+\ket{1,z_4}$ and $\ket{0,y_5}+\ket{1,z_5}$ span a 2-dimensional subspace spanned by $\ket{w_4,x_4}$ and $\ket{w_5,x_5}$.
There are two complex numbers $\a,\b$ such that $\abs{\a}^2+\abs{\b}^2=1$ and $\a(\ket{0,y_4}+\ket{1,z_4})+\b(\ket{0,y_5}+\ket{1,z_5})=\ket{w_4,x_4}$. So
\bea
\label{eq:ay0}
\a\ket{y_4}+\b\ket{y_5}={1\over\sqrt2}\ket{x_4}.
\eea
Recall that $\ket{y_4}$ and $\ket{y_5}$ both have elements of modulus $1/\sqrt6$. Applying Lemma \ref{le:la} (iii) to \eqref{eq:ay0} we obtain $\a\b=0$ or that $[\ket{y_4},\ket{y_5}]$ is equivalent to a real matrix of size $3\times2$.
The former results in the fifth product vector in $U$, and the latter
gives us a contradiction with $Y_6$ in Lemma \ref{le:mub} (vii). So both are excluded. We have proved $n\le3$. The last assertion of (i) follows from the above argument. So we have proved (i).
This completes the proof.
\epf

Based on the above results, we characterize below the four MUBs in $\bbC^6$ containing a product-vector basis. Proposition \ref{pp:prod} studies mainly the first and second MUBs, and Proposition \ref{pp:6+3+2+2} characterizes all MUBs.

\bpp
\label{pp:prod}
Suppose there are four MUBs in $\bbC^6$, the first MUB consists of six product states and the second MUB contains exactly $n$ product states. We have
\\
(i) if the first MUB and the $n$ product states are both from $\cP_1$ up to local unitaries then $n\le3$. Further if $n=3$ then up to local unitaries the first MUB is $I_3\op U$ where 
\bea
\label{eq:u130}
U={1\over\sqrt3}
\bma                    
1 & 1 & 1 \\
1 & \o& \o^2 \\
1 & \o^2 & \o \\
                     \ema
                     \cdot
                   \bma
                    1 & 0 & 0 \\
                     0 & \a & 0 \\
                     0 & 0 & \b \\
                     \ema
                     \cdot
                     {1\over\sqrt3}
\bma                    1 & 1 & 1 \\
                     1 & \o^2 & \o \\
                     1 & \o & \o^2 \\
                     \ema,
\eea
and at the same time the second MUB consists of three product states 
\bea
\label{eq:u131}
{1\over\sqrt2}\bma1\\1\ema\ox{1\over\sqrt3}\bma1\\1\\1\ema,
~~~~~~ 
{1\over\sqrt2}\bma1\\1\ema\ox{1\over\sqrt3}\bma1\\\o\\\o^2\ema,
~~~~~~
{1\over\sqrt2}\bma1\\-1\ema\ox{1\over\sqrt3}\bma1\\x\\y\ema,
\eea
and three Schmidt-rank-two entangled states 
\bea
\label{eq:u132}
u_{j0} {1\over\sqrt2}\bma1\\1\ema\ox{1\over\sqrt3}\bma1\\\o^2\\\o\ema
+
u_{j1}
{1\over\sqrt2}\bma1\\-1\ema\ox{1\over\sqrt3}\bma1\\x\o\\y\o^2\ema
+
u_{j2}
{1\over\sqrt2}\bma1\\-1\ema\ox{1\over\sqrt3}\bma1\\x\o^2\\y\o\ema,
\eea
where $\abs{\a}=\abs{\b}=\abs{x}=\abs{y}=1$, $j=0,1,2$, and $[u_{jk}]$ is an order-three unitary matrix.

(ii) If one of the first MUB and the $n$ product states in the second MUB is not from $\cP_1$ up to local unitaries  then $n\le2$.
\\
(iii) If the first MUB is $\cP_j$ for $j\in\{1,2,3\}$, then the product vector in the second MUB has the expression $(1,u)^T/\sqrt2 \ox (1,v,w)^T/\sqrt3$ where $\abs{u}=\abs{v}=\abs{w}=1$.
\\
(iv) In (i) we have $\a\ne1,\o,\o^2$, and $(x,y)\ne(\o^m,\o^n)$ for integers $m,n$. Further $[u_{jk}]$ has no zero entries.
\epp
\bpf
(i,ii,iii) The first assertion of (i), and assertion (ii) and (iii) follow from Lemma \ref{le:2prod2} and \ref{le:2prod}. Using Lemma \ref{le:2prod} we may assume that the first MUB is $I_3 \op V$, and the three product vectors in the second MUB are $\ket{0,0},\ket{0,1},\ket{1,a_0}$ up to local unitaries. Using (iii) we may assume that the first MUB and the three product vectors are respectively locally equivalent to $I_3\op W$ and ${1\over\sqrt2}\bma1\\1\ema\ox{1\over\sqrt3}\bma1\\1\\1\ema$, ${1\over\sqrt2}\bma1\\1\ema\ox{1\over\sqrt3}\bma1\\\o\\\o^2\ema$, ${1\over\sqrt2}\bma1\\-1\ema\ox{1\over\sqrt3}\bma1\\x\\y\ema$, respectively with $\abs{x}=\abs{y}=1$. Since they are from two  MUBs in $\bbC^6$, the submatrix ${1\over\sqrt3}W^\dg\bma1 & 1\\1 & \o \\1 & \o^2\ema$ have elements of modulus $1/\sqrt3$. Since the two column vectors are orthogonal, we can find an order-three complex permutation matrix $P$ such that ${1\over\sqrt3}P^\dg W^\dg\bma1 & 1\\1 & \o \\1 & \o^2\ema = {1\over\sqrt3}\bma1 & \a^*\\1 & \a^*\o \\1 & \a^*\o^2\ema$ with some $\abs{\a}=1$. So
${1\over\sqrt3}P^\dg W^\dg\bma1 & 1 & 1\\1 & \o & \o^2 \\1 & \o^2 & \o \ema = {1\over\sqrt3}\bma1 & \a^* & \b^* \\1 & \a^*\o & \b^* \o^2 \\1 & \a^*\o^2 & \b^*\o \ema$ with some $\abs{\b}=1$, because the matrices in the square brackets are unitary. Assuming $U=WP$ implies the assertion, because we can multiply any complex permutation matrix on the rhs of an MUB. So we have proved \eqref{eq:u130}-\eqref{eq:u132}.

(iv) Let $M$ be the second MUB. Since the first MUB is $I_3\op U$, we obtain that $(I_3\op U^\dg)M$ is a member of some MUB trio. Recall that $M$ contains the three product vectors in \eqref{eq:u131}, and
$U^\dg\bma1 & 1\\1 & \o \\1 & \o^2 \ema = {1\over\sqrt3}\bma1 & \a^* \\1 & \a^*\o \\1 & \a^*\o^2 \ema$. So $(I_3 \op U^\dg)M$ contains the two columns $(1,\cdots,1)/\sqrt6$ and $(1,1,1,\a^*,\a^*\o,\a^*\o^2)/\sqrt6$.
If $\a=1$, $\o$ or $\o^2$, then (by permuting the last three rows) these two columns are equivalent to two product vectors $(1,1)/\sqrt2 \ox (1,1,1)/\sqrt3$ and $(1,1)/\sqrt2 \ox (1,\o,\o^2)/\sqrt3$. It is a contradiction with the matrix $Y_7$ in Lemma \ref{le:mub} (vii). We have proved the assertion $\a\ne1,\o,\o^2$. 

It remains to prove that $(x,y)\ne(\o^m,\o^n)$ for $m,n=0,1,2$.
Note that when the assumption $n=3$ holds, the other assumption that the first MUB and the $n$ product states are both from $\cP_1$ up to local unitaries automatically holds, because we have proved assertion (ii). Then the matrix $(I_3\op U^\dg) M$ is a member of some MUB trio. If ${1\over\sqrt3}\bma1\\x\\y\ema$ in (i) is a column vector of ${1\over\sqrt3}
\bma                    
1 & 1 & 1 \\
1 & \o& \o^2 \\
1 & \o^2 & \o \\
\ema$, then the upper three rows of $(I_3\op U^\dg) M$ has a rank one matrix of size $3\times2$, or an order-three subunitary matrix. This is a contradiction with
the matrices $Y_1$ and $Y_2$ in Lemma \ref{le:mub} (vii).
Suppose ${1\over\sqrt3}\bma1\\x\\y\ema$ in (i) is a column vector of $
{1\over\sqrt3}
\bma
1 & 1 & 1 & 1 & 1 & 1\\
\o & \o^2 & 1 & \o^2 & \o & 1 \\
\o & 1 & \o^2 & \o^2 & 1 & \o\\
\ema
$. 
One can verify that
\bea
&&
\label{eq:13}
{1\over\sqrt3}
\bma                    
1 & 1 & 1 \\
1 & \o^2& \o \\
1 & \o & \o^2 \\                    
 \ema
\cdot
{1\over\sqrt3}
\bma
1 & 1 & 1 & 1 & 1 & 1\\
\o & \o^2 & 1 & \o^2 & \o & 1 \\
\o & 1 & \o^2 & \o^2 & 1 & \o\\
\ema
\notag\\
&=&
{1\over\sqrt3}
\bma
i & \o^2 i & \o^2 i & -i & -\o i & -\o i\\
\o^2 i & i & \o^2 i & -\o i & -\o i & -i \\
\o^2 i & \o^2 i & i & -\o i & -i & -\o i\\
\ema
\notag\\
&=&
{1\over\sqrt3}
\bma
1 & 1 & 1 & 1 & 1 & 1\\
\o^2 & \o & 1 & \o & 1 & \o                                                                                                                              ^2 \\
\o^2 & 1 & \o & \o & \o^2 & 1\\
\ema
\cdot
\diag(i,\o^2 i, \o^2 i, -i,-\o i, -\o i).
\eea
So ${1\over\sqrt3}
\bma                 v_1 \\
                     v_2 \\
                     v_3 \\
                     \ema:={1\over\sqrt3}
\bma                    1 & 1 & 1 \\
                     1 & \o^2 & \o \\
                     1 & \o & \o^2 \\
                     \ema \cdot 
{1\over\sqrt3}\bma1\\x\\y\ema$ is a vector in the second equation of \eqref{eq:13}.
Note that $U^\dg={1\over\sqrt3}
\bma                    
1 & 1 & 1 \\
1 & \o& \o^2 \\
1 & \o^2 & \o \\
                     \ema
                     \cdot
                   \bma
                    1 & 0 & 0 \\
                     0 & \a^* & 0 \\
                     0 & 0 & \b^* \\
                     \ema
                     \cdot
                     {1\over\sqrt3}
\bma                    1 & 1 & 1 \\
                     1 & \o^2 & \o \\
                     1 & \o & \o^2 \\
                     \ema$. Since the first MUB is $I_3\op U$ and the second MUB contains the product vector ${1\over\sqrt2}\bma1\\-1\ema\ox{1\over\sqrt3}\bma1\\x\\y\ema$, 
we obtain that $U^\dg {1\over\sqrt3}\bma1\\x
\\y\ema={1\over3}
\bma                    
1 & 1 & 1 \\
1 & \o& \o^2 \\
1 & \o^2 & \o \\
\ema
\cdot
\bma                v_1 \\
                    \a^* v_2 \\
                    \b^* v_3 \\
                     \ema$ is a vector of elements of modulus 
$1/\sqrt3$. It follows from Lemma \ref{le:mubsinkhorn} (iv) that $\bma                v_1 \\
                    \a^* v_2 \\
                    \b^* v_3 \\
                     \ema$ is proportional to one of the six column vectors in $\bma
1 & 1 & 1 & 1 & 1 & 1\\
\o & \o^2 & 1 & \o^2 & \o & 1 \\
\o & 1 & \o^2 & \o^2 & 1 & \o\\
\ema$. It follows from the third equation of \eqref{eq:13} that
$\diag(1,\a^*,\b^*)
\cdot
\bma
1 & 1 & 1 & 1 & 1 & 1\\
\o^2 & \o & 1 & \o & 1 & \o                                                                                                                              ^2 \\
\o^2 & 1 & \o & \o & \o^2 & 1\\
\ema
$ is from the columns of $\bma
1 & 1 & 1 & 1 & 1 & 1\\
\o & \o^2 & 1 & \o^2 & \o & 1 \\
\o & 1 & \o^2 & \o^2 & 1 & \o\\
\ema$. So $(\a,\b)=(\o^m,\o^n)$ with $m,n=0,1,2$. It is a contradiction with the fact that $\a\ne 1,\o,\o^2$ proved in the first part of (iv). We have proved that $(x,y)\ne(\o^m,\o^n)$ for $m,n=0,1,2$. So we obtain the observation that any order-two submatrix of the upper-right order-three submatrix of the second MUB is invertible. 

It remains to prove the last assertion of (iv). Suppose $u_{jk}=0$ for some $j,k$. Since \eqref{eq:u132} is a column vector of the second MUB, \cite[Lemma 1]{cy16} and the above observation imply that $u_{jk}=0$ for some $k'\ne k$. It is a contradiction with the fact that \eqref{eq:u132} is entangled.  
This completes the proof.
\epf

\bpp
\label{pp:6+3+2+2}
Suppose a set of four MUBs in $\bbC^6$ contains a product-vector MUB. 
\\
(i) If one of the remaining three MUBs in the set has exactly three product vectors then either of the other two MUBs has at most two product vectors. 
\\
(ii) The number of product vectors in the set is at most 6+3+2+2=13. It is achievable only if up to local unitaries, the first MUB is $I_3\op U$ with $U$ in \eqref{eq:u130}, and the remaining three MUBs respectively have the following product vectors,

\bea
\label{eq:1s20}
{1\over\sqrt2}\bma1\\1\ema\ox{1\over\sqrt3}\bma1\\1\\1\ema,
~~~~~~
{1\over\sqrt2}\bma1\\1\ema\ox{1\over\sqrt3}\bma1\\\o\\\o^2\ema,
~~~~~~
{1\over\sqrt2}\bma1\\-1\ema\ox{1\over\sqrt3}\bma1\\x\\y\ema
,
\eea
\bea
\label{eq:1s21}
{1\over\sqrt2}\bma1\\u\ema\ox{1\over\sqrt3}\bma1\\x_1\\y_1\ema,
~~~~~~
{1\over\sqrt2}\bma1\\-u\ema\ox{1\over\sqrt3}\bma1\\x_2\\y_2\ema,
\eea
\bea
\label{eq:1s22}
{1\over\sqrt2}\bma1\\v\ema\ox{1\over\sqrt3}\bma1\\x_3\\y_3\ema,
~~~~~~
{1\over\sqrt2}\bma1\\-v\ema\ox{1\over\sqrt3}\bma1\\x_4\\y_4\ema
\eea
with $\abs{x}=\abs{y}=\abs{u}=\abs{v}=\abs{x_j}=\abs{y_j}=1$.
\\
(iii) The remaining three MUBs in (ii) respectively have the following Schmidt-rank-two entangled states,
\bea
\label{eq:3a}
&&
a_{j0} {1\over\sqrt2}\bma1\\1\ema\ox{1\over\sqrt3}\bma1\\\o^2\\\o\ema
+
a_{j1}
{1\over\sqrt2}\bma1\\-1\ema\ox{1\over\sqrt3}\bma1\\x\o\\y\o^2\ema
+
a_{j2}
{1\over\sqrt2}\bma1\\-1\ema\ox{1\over\sqrt3}\bma1\\x\o^2\\y\o\ema,
\\
\label{eq:4b}
&&
b_{j0} 
{1\over\sqrt2}\bma1\\u\ema\ox{1\over\sqrt3}\bma1\\x_1\o\\y_1\o^2\ema
+
b_{j1} 
{1\over\sqrt2}\bma1\\u\ema\ox{1\over\sqrt3}\bma1\\x_1\o^2\\y_1\o\ema
+
b_{j2} 
{1\over\sqrt2}\bma1\\-u\ema\ox{1\over\sqrt3}\bma1\\x_2\o\\y_2\o^2\ema
+
b_{j3} 
{1\over\sqrt2}\bma1\\-u\ema\ox{1\over\sqrt3}\bma1\\x_2\o^2\\y_2\o\ema,
\notag\\
\eea
and
\bea
\label{eq:4c}
c_{j0} 
{1\over\sqrt2}\bma1\\v\ema\ox{1\over\sqrt3}\bma1\\x_3\o\\y_3\o^2\ema
+
c_{j1} 
{1\over\sqrt2}\bma1\\v\ema\ox{1\over\sqrt3}\bma1\\x_3\o^2\\y_3\o\ema
+
c_{j2} 
{1\over\sqrt2}\bma1\\-v\ema\ox{1\over\sqrt3}\bma1\\x_4\o\\y_4\o^2\ema
+
c_{j3} 
{1\over\sqrt2}\bma1\\-v\ema\ox{1\over\sqrt3}\bma1\\x_4\o^2\\y_4\o\ema,
\notag\\
\eea
where $[a_{jk}]$ is an order-three unitary matrix, $[b_{jk}]$ and $[c_{jk}]$ are two order-four unitary matrices. 
\epp
\bpf
(i) Using Proposition \ref{pp:prod} and local unitaries, we may assume that the first MUB is $\cP_1=I_3 \op U$ with $U$ given in \eqref{eq:u130}, and the second MUB has three product vectors in \eqref{eq:u131}, i.e., ${1\over\sqrt2}\bma1\\1\ema\ox{1\over\sqrt3}\bma1\\1\\1\ema,
$$ 
{1\over\sqrt2}\bma1\\1\ema\ox{1\over\sqrt3}\bma1\\\o\\\o^2\ema,
$ and $
{1\over\sqrt2}\bma1\\-1\ema\ox{1\over\sqrt3}\bma1\\x\\y\ema$ with $\abs{x}=\abs{y}=1$. Suppose the third MUB has three product vectors. It follows from Proposition \ref{pp:prod} (iii) that they are all of elements of modulus $1/\sqrt6$. Using Proposition \ref{pp:prod} (i) we may assume that the three product vectors are ${1\over\sqrt2}\bma1\\v\ema\ox{1\over\sqrt3}\bma1\\x_1\\y_1\ema$, ${1\over\sqrt2}\bma1\\v\ema\ox{1\over\sqrt3}\bma1\\x_2\\y_2\ema$,  and ${1\over\sqrt2}\bma1\\-v\ema\ox{1\over\sqrt3}\bma1\\x_3\\y_3\ema$, with $\abs{v}=\abs{x_j}=\abs{y_j}=1$. Thus  ${1\over\sqrt3}\bma1\\x_1\\y_1\ema$ and ${1\over\sqrt3}\bma1\\x_2\\y_2\ema$ are orthogonal. Since the vectors from the second and third MUBs are MU, the first two expressions in Eq.~\eqref{eq:u131} together with Lemma \ref{le:mubsinkhorn} (v)
imply that $v=i$ or $-i$, and ${1\over\sqrt3}\bma1\\x_1\\y_1\ema$ and ${1\over\sqrt3}\bma1\\x_2\\y_2\ema$ are orthogonal column vectors in the matrix
$
M=
{1\over\sqrt3}
\bma
1 & 1 & 1 & 1 & 1 & 1\\
\o & \o^2 & 1 & \o^2 & \o & 1 \\
\o & 1 & \o^2 & \o^2 & 1 & \o\\
\ema. 
$ The same reason implies that 
$(x,y)=(\o^m,\o^n)$ for some integers $m,n$. It is a contradiction with Proposition \ref{pp:prod} (iv), so the third MUB has at most two product vectors. If it is achievable, then the above argument implies that they are ${1\over\sqrt2}\bma1\\v\ema\ox{1\over\sqrt3}\bma1\\x_1\\y_1\ema$ and ${1\over\sqrt2}\bma1\\-v\ema\ox{1\over\sqrt3}\bma1\\x_3\\y_3\ema$ with $\abs{v}=\abs{x_j}=\abs{y_j}=1$.

(ii) follows from (i) and its proof. 
(iii) follows from (ii).
This completes the proof.
\epf

We have tried to show the claim that if a set of four MUBs in $\bbC^6$ contains a product-vector basis then any of the other three bases contains at most two product vectors. Although it is true with some specific elements $u,v,x_j,y_j$ in Proposition \ref{pp:6+3+2+2}, a proof to the claim is still missing. We believe that the expressions of MUBs in Proposition \ref{pp:6+3+2+2} will give more constraints and result in a contradiction with the existence of the four MUBs containing a product-vector basis.

\section{Conclusions}
\label{sec:con}
We have studied the corollaries of the assumption that four MUBs in $\bbC^6$ containing a product-vector basis exist. We have shown that if one of the product-vector basis and the set of product states in another MUB is not from $\cP_1$ up to local unitaries then Lemma \ref{le:mub6} holds. Otherwise the number of product states in the MUB is at most three. We also have showed that when the number is exactly three, the remaining two non-product-vector MUBs of the four MUBs in $\bbC^6$ each has at most two product vectors. We also have investigated the expressions of product states and the remaining entangled states in the MUBs. As the next step of studying the existence of four MUBs in $\bbC^6$, we may reduce the number of product vectors in four MUBs in $\bbC^6$ containing the identity matrix or a product-vector basis.

\section*{Acknowledgments}

L.C. was supported by Beijing Natural Science Foundation (4173076), the NNSF of China (Grant No. 11501024), and the Fundamental Research Funds for the Central Universities (Grant Nos. KG12001101, ZG216S1760 and ZG226S17J6). L.Y. acknowledges support from the Ministry of Science and Technology of China under Grant No. 2016YFA0301802, the funds of Hangzhou City for supporting the Hangzhou-City
Quantum Information and Quantum Optics Innovation Research Team, and the startup grant of Hangzhou Normal University.

\bibliography{channelcontrol}

\end{document}